\documentclass[]{ptptex}
\usepackage{wrapft}
\usepackage{graphicx}

\newcommand{\ket}[1]{| {#1} \rangle}     
\newcommand{\rket}[1]{| {#1} )}     
\newcommand{\rdket}[1]{|\!| {#1} )}     

\newcommand{\wtilde}[1]{\widetilde{#1}} 

\newcommand{\ovl}[1]{\overline{#1}}

\def\beq{\begin{eqnarray}}
\def\eeq{\end{eqnarray}}
\def\bsub{\begin{subequations}}
\def\esub{\end{subequations}}
\def\b{\begin{equation}}

\title{
Note on the Minimum Weight States in the $su(2)$-Algebraic Many-Fermion Model
}
\subtitle{
Extension of the Role of the Auxiliary $su(2)$-Algebra
}    

\author{
Yasuhiko {\sc Tsue},$^{1}$ 
Constan\c{c}a {\sc Provid\^encia},$^{2}$ 
Jo\~ao da {\sc Provid\^encia}$^{2}$ and 
Masatoshi {\sc Yamamura}$^{3}$  
}

\inst{
$^{1}$Physics Division, Faculty of Science, Kochi University, Kochi 780-8520, 
Japan\\
$^{2}$Departamento de F\'{i}sica, Universidade de Coimbra, 3004-516 Coimbra, 
Portugal\\
$^{3}$Department of Pure and Applied Physics, 
Faculty of Engineering Science,\\
Kansai University, Suita 564-8680, Japan
\\

}

\recdate{
\today
}

\abst{
The role of the auxiliary $su(2)$-algebra proposed by the present authors in the latest paper is 
extended. 
Its aim is to determine the minimum weight states completely. 
With the aid of scalar operators for the auxiliary algebra, 
new aspects are added to the minimum weight states in the aligned scheme 
discussed in the latest paper. 
Some simple examples are shown. 
}

\begin{document}

\maketitle


The $su(2)$-algebraic many-fermion model has played a central role 
in the field of nuclear many-body physics. 
This model consists of three generators 
${\wtilde S}_{\pm,0}$ and the orthogonal set for the irreducible 
representation is obtained by operating the raising operator ${\wtilde S}_+$ successively on 
a chosen minimum weight state $\rket{m}$. 
The state $\rket{m}$ obeys 
\beq\label{1}
{\wtilde S}_-\rket{m}=0\ , \qquad
{\wtilde S}_0\rket{m}=-s\rket{m}\ , \quad s=0,\ 1/2,\ 1,\ 3/2,\cdots .
\eeq
Recently, the present authors published a paper, in which a possible idea for systematic construction of 
$\rket{m}$ is discussed.\cite{1} 
This paper will be referred to as (I). 
Key to this problem can be found in introducing another $su(2)$-algebra, the generators of which are denoted 
as ${\wtilde R}_{\pm,0}$ in (I). 
They satisfy the relation 
\beq\label{2}
[\ {\rm any\ of}\ {\wtilde R}_{\pm,0}\ , \ {\rm any\ of}\ {\wtilde S}_{\pm,0}\ ]=0 \ .
\eeq
It is important to learn that ${\wtilde R}_{\pm,0}$ are also given in the fermion space 
giving ${\wtilde S}_{\pm,0}$. 
Dynamics under investigation is described by ${\wtilde S}_{\pm,0}$ and 
${\wtilde R}_{\pm,0}$ play a supporting role in the determination of 
$\rket{m}$. 
In view of this role, in (I), we called this algebra as the auxiliary $su(2)$-algebra.

For the convenience of later discussion, first, main part of (I) is recapitulated briefly. 
We treat many-fermion system confined in $4\Omega_0$ single-particle states. 
Here, $\Omega_0$ denotes integer or half-integer. 
Since $4\Omega_0$ is an even number, all single-particle states are divided into 
equal parts $P$ and ${\ovl P}$. 
Therefore, as a partner, each single-particle state belonging to $P$ can find a single-particle state in ${\ovl P}$. 
We express the partner of the state $\alpha$ belonging to $P$ as ${\bar \alpha}$ and 
fermion operators in $\alpha$ and ${\bar \alpha}$ are denoted as 
$({\tilde c}_\alpha,{\tilde c}_{\alpha}^*)$ and $({\tilde c}_{\bar \alpha}, {\tilde c}_{\bar \alpha}^*)$, 
respectively. 
As the generators ${\wtilde S}_{\pm,0}$, we adopt the following form:
\beq\label{3}
{\wtilde S}_+=\sum_{\alpha}s_{\alpha}{\tilde c}_{\alpha}^*{\tilde c}_{\bar \alpha}^*\ , \qquad
{\wtilde S}_-=\sum_{\alpha}s_{\alpha}{\tilde c}_{\bar \alpha}{\tilde c}_{\alpha}\ , \qquad
{\wtilde S}_0=\frac{1}{2}\sum_{\alpha}
({\tilde c}_{\alpha}^*{\tilde c}_{\alpha}+{\tilde c}_{\bar \alpha}^*{\tilde c}_{\bar \alpha})-\Omega_0 \ .
\eeq
The symbol $s_{\alpha}$ denotes real number satisfying $s_{\alpha}^2=1$. 
The sum $\sum_{\alpha}\ (\sum_{\bar \alpha})$ is carried in all single-particle states in 
$P$ (${\ovl P})$ and, 
then, 
we have $\sum_{\alpha}1=2\Omega_0\ (\sum_{\bar \alpha}1=2\Omega_0)$. 
On the other hand, ${\wtilde R}_{\pm,0}$ are defined in the form 
\beq\label{4}
{\wtilde R}_+=\sum_{\alpha}{\tilde c}_{\alpha}^*{\tilde c}_{\bar \alpha}\ , \qquad
{\wtilde R}_-=\sum_{\alpha}{\tilde c}_{\bar \alpha}^*{\tilde c}_{\alpha}\ , \qquad
{\wtilde R}_0=\frac{1}{2}\sum_{\alpha}
({\tilde c}_{\alpha}^*{\tilde c}_{\alpha}-{\tilde c}_{\bar \alpha}^*{\tilde c}_{\bar \alpha}) \ .
\eeq
It is easily verified that the operators (\ref{3}) and (\ref{4}) satisfy the 
relation (\ref{2}). 
In (I), we called the $su(2)$-algebras $({\wtilde S}_{\pm,0})$ and $({\wtilde R}_{\pm,0})$ as 
the $S$- and the $R$-spin, respectively.

The idea for the determination of $\rket{m}$ presented in (I) is summarized as follows: 
First, for the determination of the minimum weight state, $\rket{m_0}$, for the $R$-spin, we set up the 
relation
\beq\label{5}
{\wtilde R}_-\rket{m_0}=0\ , \qquad
{\wtilde R}_0\rket{m_0}=-r\rket{m_0}\ , \quad
r=0,\ 1/2,\ 1,\ 3/2,\cdots .
\eeq
Then, by operating the raising operator ${\wtilde R}_+$ successively on $\rket{m_0}$, the orthogonal set for the $R$-spin 
is obtained as 
\beq\label{6}
\rket{m}=\left({\wtilde R}_+\right)^\lambda \rket{m_0}\ .
\eeq
The state $\rket{m}$ satisfies the relation (\ref{1}), if $\rket{m_0}$ satisfies 
\beq\label{7}
{\wtilde S}_-\rket{m_0}=0\ , \qquad
{\wtilde S}_0\rket{m_0}=-s\rket{m_0}\ .
\eeq
Here, we used the relation (\ref{2}). 
The above is our basic idea shown in (I).

In (I), we adopted the following form for $\rket{m_0}$: 
\beq\label{8}
\rket{m_0}=
\left\{
\begin{array}{ll}
\rket{0}\ , & (r=0) \\
\displaystyle \prod_{i=1}^{2r}{\tilde c}_{{\bar \alpha}_i}^*\rket{0}\ (=\rket{r;({\bar \alpha})})\ . &
(r=1/2,\ 1,\ 3/2\, \cdots )
\end{array}
\right.
\eeq
Here, $({\bar \alpha})$ denotes the set $({\bar \alpha}_1, {\bar \alpha}_2, \cdots ,{\bar \alpha}_{2r})$. 
Beforehand, the rule of the arrangement of $({\bar \alpha})$ should be provided. 
For choosing $({\bar \alpha})$ for a given value of $r$, there exist 
$(2\Omega_0)!/(2r)!(2\Omega_0-2r)!$ possibilities. 
They form the orthogonal set: 
\beq\label{9}
(r;({\bar \alpha})\rket{r';({\bar \alpha}')}=
\left\{
\begin{array}{ll}
\displaystyle \prod_{i=1}^{2r}\delta_{{\bar \alpha}_i{\bar \alpha}'_i} \ , & (r=r') \\
0 \ . & (r\neq r') \\
\end{array}
\right.
\eeq
For the state $\rket{r;({\bar \alpha})}$, we have 
\bsub\label{10}
\beq
& &{\wtilde R}_-\rket{r;({\bar \alpha})}=0\ , \qquad
{\wtilde R}_0\rket{r;({\bar \alpha})}=-r\rket{r;({\bar \alpha})} \ , 
\label{10a}\\
& &{\wtilde S}_-\rket{r;({\bar \alpha})}=0\ , \qquad
{\wtilde S}_0\rket{r;({\bar \alpha})}=-s\rket{r;({\bar \alpha})} \ . 
\qquad (s=\Omega_0-r) 
\label{10b}
\eeq
\esub
With the use of the above result, the relation (\ref{6}) leads us to 
\beq\label{11}
\rket{m}=
\left({\wtilde R}_+\right)^{r+r_0}\rket{m_0}
=\left({\wtilde R}_+\right)^{r+r_0}\rket{r;({\bar \alpha})}
=\rket{rr_0;({\bar \alpha})}\ .
\eeq
Clearly, $\rket{m}$ is the minimum weight state of the $S$-spin and the 
following relation is derived: 
\beq\label{12}
r+s=\Omega_0 \ .
\eeq
It should be stressed that the magnitudes of the $R$- and the $S$-spin are 
restricted by the relation (\ref{12}), which is derived in the frame of the state (\ref{8}) 
and $2r$ denotes the seniority number. 
In (I), we called the state (\ref{8}) as the minimum weight state in the 
aligned scheme. 
Concerning the form (\ref{8}), we mentioned in the concluding remarks of (I) (\S 8) 
that, by showing an instructive example (two-fermion states), 
the form (\ref{8}) does not cover 
all cases. 
It covers only the aligned scheme. 
In this paper, various states appear and we omit numerical factor related to the
normalization constant of any state.

Main aim of this paper is to present an extended form of $\rket{m_0}$ 
from the form in the aligned scheme (\ref{8}) 
and the basic idea comes from the extension of the role of the $R$-spin, that is, 
the introduction of scalar operators for the $R$-spin.  
For this aim, we will make a preparatory argument. 
Let any operator in the present many-fermion space be denoted as ${\wtilde P}^*$. 
Generally, we have the relation 
\beq\label{13}
{\wtilde S}_-{\wtilde P}^*\rket{r;({\bar \alpha})}\neq 0 \ . \qquad
\left({\wtilde S}_-\rket{r;({\bar \alpha})}=0\right)
\eeq
The relation (\ref{13}) indicates that ${\wtilde P}^*\rket{r;({\bar \alpha})}$ is not 
minimum weight state for the $S$-spin. 
Then, let us define ${\wtilde {\cal P}}^*$, the counterpart of ${\wtilde P}^*$, in the form 
\beq\label{14}
{\wtilde {\cal P}}^*=-{\wtilde P}^*{\wtilde S}_0
-\frac{1}{2}[\ {\wtilde S}_-\ , \ {\wtilde P}^*\ ]{\wtilde S}_+ \ . 
\eeq
The operator ${\wtilde {\cal P}}^*$ satisfies the relation 
\beq\label{15}
[\ {\wtilde S}_-\ , \ {\wtilde {\cal P}}^*\ ]={\wtilde P}^*{\wtilde S}_-\ , 
\eeq
if ${\wtilde P}^*$ obeys the condition 
\beq\label{16}
[\ {\wtilde S}_-\ , \ [\ {\wtilde S}_-\ , \ {\wtilde P}^*\ ] \ ]=0\ .
\eeq
The relation (\ref{15}) leads us to 
\beq\label{17}
{\wtilde S}_-{\wtilde {\cal P}}^*\rket{r;({\bar \alpha})}=0 \ .
\eeq
The above argument tells us that we can construct ${\wtilde {\cal P}}^*$, 
the counterpart of ${\wtilde P}^*$ which obeys the relation (\ref{16}), 
and ${\wtilde {\cal P}}^*\rket{r;({\bar \alpha})}$ is also 
the minimum weight state of the $R$-spin. 
If ${\wtilde P}^*$ is a scalar operator for the $R$-spin, 
${\wtilde {\cal P}}^*$ is also scalar:
\beq\label{19}
{\rm if}\quad [\ {\wtilde R}_{\pm,0}\ , \ {\wtilde P}^*\ ]=0\ , \qquad
 [\ {\wtilde R}_{\pm,0}\ , \ {\wtilde {\cal P}}^*\ ]=0 \ . 
\eeq
It is also noted that in the case of the product $\prod_l{\wtilde P}_l^*$, we have 
\beq\label{20}
{\wtilde S}_-\prod_{l}{\wtilde {\cal P}}^*_l\rket{r;({\bar \alpha})}=0 \ .
\eeq
Of course, ${\wtilde {\cal P}}_l^*$ is the operator derived from ${\wtilde P}_l^*$ through the relation (\ref{14}). 
For example, in the case of the product ${\wtilde {\cal P}}_1^*{\wtilde {\cal P}}_2^*$, 
we have 
\beq\label{20-2}
[\ {\wtilde S}_-\ , \ {\wtilde {\cal P}}_1^*{\wtilde {\cal P}}_2^*\ ]=
\left({\wtilde P}_1^*{\wtilde P}_2^*+{\wtilde P}_1^*{\wtilde {\cal P}}_2^*
+{\wtilde {\cal P}}_1^*{\wtilde P}_2^*\right){\wtilde S}_- \ .
\eeq
The above is the preparatory argument.

The above argument suggests us that our next task is to search concrete forms of the scalar operators. 
We note that the set $({\tilde c}_\alpha^*,{\tilde c}_{\bar \alpha}^*)$ composes a spinor, i.e., a tensor operator 
of rank 1/2: 
\bsub\label{21}
\beq
& &[\ {\wtilde R}_+\ , \ {\tilde c}_\alpha^*\ ]=0\ , \qquad
[\ {\wtilde R}_-\ , \ {\tilde c}_\alpha^*\ ]={\tilde c}_{\bar \alpha}^*\ , \qquad
[\ {\wtilde R}_0\ , \ {\tilde c}_\alpha^*\ ]=+\frac{1}{2}{\tilde c}_{\alpha}^*\ , 
\label{21a}\\
& &[\ {\wtilde R}_-\ , \ {\tilde c}_{\bar \alpha}^*\ ]=0\ , \qquad
[\ {\wtilde R}_+\ , \ {\tilde c}_{\bar \alpha}^*\ ]={\tilde c}_{\alpha}^*\ , \qquad
[\ {\wtilde R}_0\ , \ {\tilde c}_{\bar \alpha}^*\ ]=-\frac{1}{2}{\tilde c}_{\bar \alpha}^*\ . 
\label{21b}
\eeq
\esub
By using the above spinors, 
the following scalar operators can be constructed: 
\bsub\label{22}
\beq
& &{\wtilde S}_{\mu\nu}^*={\tilde c}_{\mu}^*{\tilde c}_{\bar \nu}^*
-{\tilde c}_{\bar \mu}^*{\tilde c}_{\nu}^*\ , \qquad ({\wtilde S}_{\mu\nu}^*={\wtilde S}_{\nu\mu}^*\ , \ \mu\neq \nu) 
\label{22a}\\
& &{\wtilde S}_\lambda^*={\tilde c}_{\lambda}^*{\tilde c}_{\bar \lambda}^*\ , 
\qquad ({\wtilde S}_{\mu\mu}^*=2{\wtilde S}_\mu^*)
\label{22b}
\eeq
\esub
The operators ${\wtilde S}_{\mu}^*$ is a special case of ${\wtilde S}_{\mu\nu}^*\ (\mu=\nu)$. 
But, from the reason mentioned below, it may be convenient to discriminate between both forms for the treating. 
The operators (\ref{22a}) and (\ref{22b}) have the properties 
\beq\label{23}
{\wtilde S}_{\lambda\mu}^*\cdot{\wtilde S}_{\lambda\nu}^*=-{\wtilde S}_{\lambda}^*\cdot{\wtilde S}_{\mu\nu}^*\ , \qquad
\left({\wtilde S}_{\mu\nu}^*\right)^2=-2{\wtilde S}_{\mu}^*\cdot{\wtilde S}_{\nu}^*\ .
\eeq
Further, the operator (\ref{22a}) satisfies the relation 
\beq\label{24}
{\wtilde S}_{\mu\alpha_i}^*\rket{r;{\bar \alpha}_1\cdots{\bar \alpha}_{i-1}{\bar \alpha}_i{\bar \alpha}_{i+1}\cdots{\bar \alpha}_{2r}}
=(-)^{\phi_i}{\wtilde S}_{\alpha_i}^*\rket{r;{\bar \alpha}_1\cdots{\bar \alpha}_{i-1}{\bar \mu}{\bar \alpha}_{i+1}\cdots{\bar \alpha}_{2r}}\ .
\eeq
Here, $(-)^{\phi_i}$ denotes the phase factor coming from the anti-commutation relation. 
The relation (\ref{24}) teaches us that the operation of ${\wtilde S}_{\mu\nu}^*$ on 
$\rket{r;({\bar \alpha})}$ changes the minimum weight state in the aligned scheme to another, if the 
subscript $\mu$ or $\nu$ is identical to any of $(\alpha_1,\cdots,\alpha_{2r})$. 
In order to make the minimum weight state in the aligned scheme unchange, we require that on the occasion of 
operating on $\rket{r;({\bar \alpha})}$, 
the subscripts of any of ${\wtilde S}_{\mu\nu}^*$ are all different from any of 
$(\alpha_1,\cdots,\alpha_{2r})$. 
Further, the relation (\ref{23}) makes us require that on the occasion of successive operation of 
${\wtilde S}_{\mu\nu}^*$ etc., all the subscripts are different from one another. 
If this requirement is permitted, the operation of ${\wtilde S}_{\mu\nu}^*$ can be performed 
independently of the operation of ${\wtilde S}_{\lambda}^*$. 
In other words, we can forget the composite nature of ${\wtilde S}_{\lambda}^*$ and ${\wtilde S}_{\mu\nu}^*$.

Next, we consider the case of the tensor of rank 1, ${\wtilde Q}_{\alpha\beta}^{*(1)}(\kappa)$ 
$(\kappa=+1,0,-1)$, 
which consist of the products of two fermion creation operators. 
The operators ${\wtilde Q}_{\alpha\beta}^{*(1)}(\kappa)$ are given in the form 
\beq\label{25}
{\wtilde Q}_{\alpha\beta}^{*(1)}(+1)={\tilde c}_{\alpha}^*{\tilde c}_{\beta}^*\ , \qquad
{\wtilde Q}_{\alpha\beta}^{*(1)}(0)=\frac{1}{\sqrt{2}}({\tilde c}_{\alpha}^*{\tilde c}_{\bar \beta}^*
+{\tilde c}_{\bar \alpha}^*{\tilde c}_{\beta}^*)\ , \qquad
{\wtilde Q}_{\alpha\beta}^{*(1)}(-1)={\tilde c}_{\bar \alpha}^*{\tilde c}_{\bar \beta}^*\ . \qquad
\eeq
We can construct the scalar from the tensor (\ref{25}) in the form 
\bsub\label{26}
\beq\label{26a}
{\wtilde P}_{\alpha\beta,\lambda\mu}^{*(1)}=\sum_{\kappa}(-)^{1-\kappa}
{\wtilde Q}_{\alpha\beta}^{*(1)}(\kappa){\wtilde Q}_{\lambda\mu}^{*(1})(-\kappa) \ . 
\eeq
The operator ${\wtilde P}_{\alpha\beta,\lambda\mu}^{*(1)}$ can be rewritten as 
\beq\label{27}
{\wtilde P}_{\alpha\beta,\lambda\mu}^{*(1)}=-\frac{1}{2}\left(
{\wtilde S}_{\alpha\lambda}^*{\wtilde S}_{\beta\mu}^*-{\wtilde S}_{\alpha\mu}^*
{\wtilde S}_{\beta\lambda}^*\right) \ .
\eeq
\esub
It is noted that ${\wtilde S}_{\alpha\lambda}^*$ etc. in the form (\ref{27}) contain the cases $\alpha=\lambda$ etc. 
which reduce to $2{\wtilde S}_\alpha^*$ etc. 
Further, we have the following form for the tensor of rank 3/2, 
which consist of the products of three fermion creation operators:
\beq\label{28}
& &{\wtilde Q}_{\alpha\beta\gamma}^{*(3/2)}(+3/2)=
{\tilde c}_\alpha^*{\tilde c}_{\beta}^*{\tilde c}_{\gamma}^*\ , \quad
{\wtilde Q}_{\alpha\beta\gamma}^{*(3/2)}(1/2)=
\frac{1}{\sqrt{3}}({\tilde c}_{\alpha}^*{\tilde c}_{\beta}^*{\tilde c}_{\bar \gamma}^*
+{\tilde c}_{\alpha}^*{\tilde c}_{\bar \beta}^*{\tilde c}_{\gamma}^*
+{\tilde c}_{\bar \alpha}^*{\tilde c}_{\beta}^*{\tilde c}_{\gamma}^*) \ , \nonumber\\
& &{\wtilde Q}_{\alpha\beta\gamma}^{*(3/2)}(-1/2)=
\frac{1}{\sqrt{3}}({\tilde c}_{\bar \alpha}^*{\tilde c}_{\bar \beta}^*{\tilde c}_{\gamma}^*
+{\tilde c}_{\bar \alpha}^*{\tilde c}_{\beta}^*{\tilde c}_{\bar \gamma}^*
+{\tilde c}_{\alpha}^*{\tilde c}_{\bar \beta}^*{\tilde c}_{\bar \gamma}^*) \ , \quad
{\wtilde Q}_{\alpha\beta\gamma}^{*(3/2)}(-3/2)={\tilde c}_{\bar \alpha}^*{\tilde c}_{\bar \beta}^*{\tilde c}_{\bar \gamma}^*\ .
\quad
\eeq
The scalar operator constructed from ${\wtilde Q}_{\alpha\beta\gamma}^{*(3/2)}(\kappa)$ can be 
expressed as 
\beq\label{29}
{\wtilde P}_{\alpha\beta\gamma,\lambda\mu\nu}^{*(3/2)}&=&
\sum_{\kappa}(-)^{3/2-\kappa}{\wtilde Q}_{\alpha\beta\gamma}^{*(3/2)}(\kappa)
{\wtilde Q}_{\lambda\mu\nu}^{*(3/2)}(-\kappa)\
\nonumber\\
&=&
-\frac{1}{3}\left({\wtilde S}_{\alpha\lambda}^*{\wtilde S}_{\beta\mu}^*{\wtilde S}_{\gamma\nu}^*
+{\wtilde S}_{\alpha\mu}^*{\wtilde S}_{\beta\nu}^*{\wtilde S}_{\gamma\lambda}^*
+{\wtilde S}_{\alpha\nu}^*{\wtilde S}_{\beta\lambda}^*{\wtilde S}_{\gamma\mu}^*\right) \ .
\eeq
In the relation (\ref{29}), the cases $\alpha=\lambda$ etc. are also contained. 
The relations (\ref{27}) and (\ref{29}) tell us that the scalar 
operators constructed by the tensors ${\wtilde Q}_{\alpha\beta}^{*(1)}(\kappa)$ and 
${\wtilde Q}_{\alpha\beta\gamma}^{*(3/2)}(\kappa)$ can be expressed in terms of 
${\wtilde S}_{\lambda}^*$ and ${\wtilde S}_{\mu\nu}^*$ defined in the relation (\ref{22}). 
Judging from the above two cases, probably, the scalar operators 
constructed from the tensors of 
any rank consisting of the products of any number of fermion creation operators 
may be expressed in terms of 
${\wtilde S}_{\lambda}^*$ and ${\wtilde S}_{\mu\nu}^*$. 
For example, we have the following case: 
\bsub\label{29add}
\beq
& &{\wtilde P}_{\alpha\beta\gamma,\lambda\mu\nu}^{*(1/2)}
=\sum_{\kappa}(-)^{\frac{1}{2}-\kappa}{\wtilde Q}_{\alpha\beta\gamma}^{*(1/2)}(\kappa){\wtilde Q}_{\lambda\mu\nu}^{*(1/2)}(-\kappa)
\nonumber\\
& &\qquad\qquad
=\frac{1}{6}\biggl(
{\wtilde S}_{\alpha\lambda}^*{\wtilde S}_{\beta\gamma}^*{\wtilde S}_{\mu\nu}^*-
{\wtilde S}_{\alpha\gamma}^*{\wtilde S}_{\beta\lambda}^*{\wtilde S}_{\mu\nu}^*
-
{\wtilde S}_{\alpha\mu}^*{\wtilde S}_{\beta\gamma}^*{\wtilde S}_{\lambda\nu}^*
+{\wtilde S}_{\alpha\gamma}^*{\wtilde S}_{\beta\mu}^*{\wtilde S}_{\lambda\nu}^*
\biggl)\ , 
\label{adda}\\
& &{\wtilde Q}_{\alpha\beta\gamma}^{*(1/2)}(\kappa=+1/2)
=
\langle 1 1  1\!/2 -\!\!1\!/2\ket{1\!/2\ 1\!/2}{\wtilde Q}_{\alpha\beta}^{*(1)}(+1){\tilde c}_{\bar \gamma}^*
\nonumber\\
& &\qquad\qquad\qquad\qquad\ \ \ \ \ 
+\langle 1 0 1\!/2 1\!/2\ket{1/2\ 1/2}{\wtilde Q}_{\alpha\beta}^{*(1)}(0){\tilde c}_{\gamma}^*
\nonumber\\
& &\qquad\qquad\qquad\qquad\ 
=
\frac{1}{\sqrt{6}}\biggl(
{\tilde c}_{\alpha}^*{\wtilde S}_{\beta\gamma}^*-{\tilde c}_{\beta}^*{\wtilde S}_{\alpha\gamma}^*\biggl)\ , 
\label{addb}\\
& &{\wtilde Q}_{\alpha\beta\gamma}^{*(1/2)}(\kappa=-1/2)
=
\langle 1 -\!\!1  1\!/2\ 1\!/2\ket{1\!/2\ -\!\!1\!/2}{\wtilde Q}_{\alpha\beta}^{*(1)}(-1){\tilde c}_{\gamma}^*
\nonumber\\
& &\qquad\qquad\qquad\qquad\ \ \ \ \ 
+\langle 1 0 1\!/2 -\!\!1\!/2\ket{1/2 -\!\!1/2}{\wtilde Q}_{\alpha\beta}^{*(1)}(0){\tilde c}_{\bar \gamma}^*
\nonumber\\
& &\qquad\qquad\qquad\qquad\ 
=
\frac{1}{\sqrt{6}}\biggl(
{\tilde c}_{\bar \alpha}^*{\wtilde S}_{\beta\gamma}^*-{\tilde c}_{\bar \beta}^*{\wtilde S}_{\alpha\gamma}^*\biggl)\ . 
\label{addc}
\eeq
\esub
Here, $\langle 1\pm\!\!1\ 1/2\mp\!\!1/2\ket{1/2\pm\!\!1/2}$ and 
$\langle 1 0\ 1/2 \ \pm\!\!1/2\ket{1/2\pm\!\!1/2}$ denote the Clebsch-Gordan coefficients.

With the use of the scalar operators ${\wtilde S}_{\lambda}^*$ and ${\wtilde S}_{\mu\nu}^*$, 
we define the following state: 
\beq
& &\rket{p;(\lambda),q;(\mu\nu),r;({\bar \alpha})}
={\wtilde P}^*(p;(\lambda),q;(\mu\nu))\rket{r;({\bar \alpha})}\ , 
\label{29-2}\\
& &{\wtilde P}^*(p;(\lambda),q;(\mu\nu))=\prod_{k=1}^p {\wtilde S}_{\lambda_k}^*\prod_{j=1}^q{\wtilde S}_{\mu_j\nu_j}^*\ . 
\label{30}
\eeq
If any of $(\mu_1,\nu_1),\cdots ,\ (\mu_q,\nu_q)$ is different from 
any other of $(\mu_1,\nu_1), \cdots ,\ (\mu_q, \nu_q)$ and also different from any of $\alpha_1,\cdots \, \alpha_{2r}$, 
we can treat ${\wtilde S}_{\lambda}^*$ and ${\wtilde S}_{\mu\nu}^*$ independently of one another. 
The set of the states (\ref{29-2}) forms an orthogonal set. 
It can be understood from the following: 
The state (\ref{29-2}) is an eigenstate of $2\Omega_0$ fermion number operators with the 
eigenvalues 0, 1 and 2, where the fermion number operator in the states $\alpha$ and 
${\bar \alpha}$ is given as 
\beq\label{31}
{\wtilde N}_{\alpha}={\tilde c}_{\alpha}^*{\tilde c}_{\alpha}
+{\tilde c}_{\bar \alpha}^*{\tilde c}_{\bar \alpha}\ . 
\eeq
Any set of the eigenvalues is different from any other one. 
The state (\ref{29-2}) satisfies the relation 
\bsub\label{32}
\beq
& &{\wtilde R}_-\rket{p;(\lambda),q;(\mu\nu),r;({\bar \alpha})}=0 \ , 
\label{32a}\\
& &{\wtilde R}_0\rket{p;(\lambda),q;(\mu\nu),r;({\bar \alpha})}
=-r\rket{p;(\lambda),q;(\mu\nu),r;({\bar \alpha})} \ , 
\label{32b} 
\eeq
\vspace{-0.8cm}
\esub
\bsub\label{33}
\beq
& &{\wtilde N}\rket{p;(\lambda),q;(\mu\nu),r;({\bar \alpha})}
=n\rket{p;(\lambda),q;(\mu\nu),r;({\bar \alpha})} \ \quad 
\label{33a}\\ 
& &{\wtilde N}=\sum_{\alpha}{\wtilde N}_{\alpha}\ , \qquad
n=2(p+q+r)\ . 
\label{33b}
\eeq
\esub
However, the state (\ref{29-2}) is not minimum weight state for $({\wtilde S}_{\pm,0})$. 
Then, replacing ${\wtilde P}^*$ with ${\wtilde P}^*(p;(\lambda),q;(\mu\nu))$ in the 
relation (\ref{15}), we obtain ${\wtilde {\cal P}}^*(p;(\lambda),q;(\mu\nu))$ and 
we define the state 
\beq\label{34}
\rdket{p;(\lambda),q;(\mu\nu),r;({\bar \alpha})}={\wtilde {\cal P}}^*(p;(\lambda),q;(\mu\nu))\rket{r;({\bar \alpha})} \ . 
\eeq
The state (\ref{34}) is the minimum weight state of the 
$R$-spin and also the $S$-spin, i.e., $\rket{m_0}$. 
With the aid of the successive operation of ${\wtilde R}_+$, we obtain the minimum weight state 
for $({\wtilde S}_{\pm,0})$ in the form 
\beq\label{35}
\rdket{p;(\lambda),q;(\mu\nu),rr_0;({\bar \alpha})}
&=&\left({\wtilde R}_+\right)^{r+r_0}
\rdket{p;(\lambda),q;(\mu\nu),r;({\bar \alpha})} \nonumber\\
&=&{\wtilde {\cal P}}^*(p;(\lambda),q;(\mu\nu))\rket{rr_0;({\bar \alpha})} \ , 
\eeq
\vspace{-0.8cm}
\bsub\label{36}
\beq
& &{\wtilde S}_-\rdket{p;(\lambda),q;(\mu\nu),rr_0;({\bar \alpha})}=0\ , 
\label{36a}\\
& &{\wtilde S}_0\rdket{p;(\lambda),q;(\mu\nu),rr_0;({\bar \alpha})} 
=-s\rdket{p;(\lambda),q;(\mu\nu),rr_0;({\bar \alpha})}\ . \ \ \ \ 
\label{36b}
\eeq
\esub
Here, we have 
\beq\label{37}
p+q+r+s=\Omega_0\ . 
\eeq
The quantity $2(p+q+r)$ denotes the seniority number: 
\beq\label{38}
n=2(p+q+r)\ .
\eeq
Of course, in the case $p=q=0$, the expression (\ref{37}) is reduced to the previous result (\ref{12}), 
which was derived in (I). 
We already mentioned that the set of the states ({\ref{29-2}) forms 
the orthogonal set. 
But, generally, it may be very hard to show that the set of the states (\ref{34}) forms the orthogonal set or not.

First, we pay an attention to the following: 
The operators ${\tilde c}_{\alpha}^*$ and ${\tilde c}_{\bar \alpha}^*$ 
satisfy the condition (\ref{16}). 
The counterpart of ${\tilde c}_{\alpha}^*$ and ${\tilde c}_{\bar \alpha}^*$, 
which are denoted by ${\tilde {\cal C}}_{\alpha}^*$ and ${\tilde {\cal C}}_{\bar \alpha}^*$, 
respectively, are obtained by replacing ${\wtilde P}^*$ with 
${\tilde c}_{\alpha}^*$ and ${\tilde c}_{\bar \alpha}^*$ in the relation (\ref{14}) for ${\wtilde {\cal P}}^*$: 
\bsub\label{40-0}
\beq
& &{\tilde {\cal C}}_{\alpha}^*=-{\tilde c}_{\alpha}^*{\wtilde S}_0
-\frac{1}{2}(s_{\alpha}{\tilde c}_{\bar \alpha}){\wtilde S}_+\ , 
\label{40-0a}\\
& &{\tilde {\cal C}}_{\bar \alpha}^*=-{\tilde c}_{\bar \alpha}^*{\wtilde S}_0
+\frac{1}{2}(s_{\alpha}{\tilde c}_{\alpha}){\wtilde S}_+\ . 
\label{40-0b}
\eeq
\esub
It is verified that the anti-commutators for all combinations of the operators (\ref{40-0}) vanish. 
Therefore, the anti-symmetric property of any state constructed by the operators (\ref{40-0}) 
may be guaranteed. 
But, the explicit use of the anti-commutation relations for the operators 
(\ref{40-0}) and their hermitian conjugate may be ineffective, because their forms are complicated. 
With the use of the form (\ref{40-0}), we get 
\beq\label{41-0}
{\tilde {\cal C}}_{\alpha}^*{\tilde {\cal C}}_{\bar \beta}^*
-{\tilde {\cal C}}_{\bar \alpha}^*{\tilde {\cal C}}_{\beta}^*
&=&
({\tilde c}_{\alpha}^*{\tilde c}_{\bar \beta}^*-{\tilde c}_{\bar \alpha}^*{\tilde c}_{\beta}^*)
{\wtilde S}_0\left({\wtilde S}_0-\frac{1}{2}\right)
\nonumber\\
& &-\frac{1}{2}{\wtilde S}_+\left({\wtilde S}_0-\frac{1}{2}\right)
\left[
s_{\beta}({\tilde c}_{\alpha}^*{\tilde c}_{\beta}+{\tilde c}_{\bar \alpha}^*{\tilde c}_{\bar \beta}-\delta_{\alpha\beta})
+s_{\alpha}({\tilde c}_{\beta}^*{\tilde c}_{\alpha}+{\tilde c}_{\bar \beta}^*{\tilde c}_{\bar \alpha}-\delta_{\alpha\beta})
\right]
\nonumber\\
& &
-\frac{1}{4}\left({\wtilde S}_+\right)^2s_{\alpha}s_{\beta}
({\tilde c}_{\bar \beta}{\tilde c}_{\alpha}-{\tilde c}_{\beta}{\tilde c}_{\bar \alpha})\ .
\eeq
We define the counterparts of ${\wtilde S}_{\mu\nu}^*$ and 
${\wtilde S}_{\lambda}^*$, ${\wtilde {\cal S}}_{\mu\nu}^*$ and 
${\wtilde {\cal S}}_{\lambda}^*$, in the form 
\bsub\label{42-0}
\beq
& &{\wtilde {\cal S}}_{\mu\nu}^*={\tilde {\cal C}}_{\mu}^*{\tilde {\cal C}}_{\bar \nu}^*
-{\tilde {\cal C}}_{\bar \mu}^*{\tilde {\cal C}}_{\nu}^*\ , \quad (\mu\neq \nu)
\label{42-0a}\\
& &{\wtilde {\cal S}}_{\lambda}^*=\frac{1}{2}\left(
{\tilde {\cal C}}_{\lambda}^*{\tilde {\cal C}}_{\bar \lambda}^*
-{\tilde {\cal C}}_{\bar \lambda}^*{\tilde {\cal C}}_{\lambda}^*\right)\ .
\label{42-0b}
\eeq
\esub
Then, the form (\ref{41-0}) leads us to the following:
\bsub\label{43-0}
\beq
{\wtilde {\cal S}}_{\mu\nu}^*
&=&
{\wtilde S}_{\mu\nu}^*\left({\wtilde S}_0-\frac{1}{2}\right){\wtilde S}_0
\nonumber\\
& &
-\frac{1}{2}{\wtilde S}_+\left({\wtilde S}_0-\frac{1}{2}\right)
\left[
s_{\nu}({\tilde c}_{\mu}^*{\tilde c}_{\nu}+{\tilde c}_{\bar \mu}^*{\tilde c}_{\bar \nu})
+s_{\mu}({\tilde c}_{\nu}^*{\tilde c}_{\mu}+{\tilde c}_{\bar \nu}^*{\tilde c}_{\bar \mu})
\right]
\nonumber\\
& &
-\frac{1}{4}\left({\wtilde S}_+\right)^2s_{\mu}s_{\nu}{\wtilde S}_{\mu\nu}\ , 
\label{43-0a}\\
{\wtilde {\cal S}}_{\lambda}^*&=&
{\wtilde S}_{\lambda}^*\left({\wtilde S}_0-\frac{1}{2}\right){\wtilde S}_0
-{\wtilde S}_+\left({\wtilde S}_0-\frac{1}{2}\right)\frac{s_{\lambda}}{2}\left({\wtilde N}_{\lambda}-1\right)
-\frac{1}{4}\left({\wtilde S}_+\right)^2{\wtilde S}_{\lambda}\ . 
\label{43-0b}
\eeq
\esub
First, we investigate ${\wtilde {\cal S}}_{\mu\nu}^*$. 
By operating $\prod_{j=1}^q{\wtilde {\cal S}}_{\mu_j\nu_j}^*$ on 
$\rket{r;({\bar \alpha})}$, we have the following form:
\beq\label{40}
\prod_{j=1}^q{\wtilde {\cal S}}_{\mu_j\nu_j}^*\rket{r;({\bar \alpha})}=
\frac{(\Omega_0-r+q)!}{(\Omega_0-r)!}\cdot
\prod_{j=1}^q{\wtilde S}_{\mu_j\nu_j}^*\rket{r;({\bar \alpha})}\ .
\eeq
For the derivation of the form (\ref{40}), we used the requirement that any of $\mu$, ${\bar \mu}$, $\nu$ 
and ${\bar \nu}$ appearing on the second term in the expression (\ref{43-0a}) does not 
appear in the state $\rket{r;({\bar \alpha})}$. 
The relation (\ref{40}) tells us that practically we can treat ${\wtilde {\cal S}}_{\mu\nu}^*$ by 
regarding it as ${\wtilde S}_{\mu\nu}^*$ and the factor $(\Omega_0-r)!$ gives us an inequality $r \leq \Omega_0$. 
Next, we consider the relation (\ref{43-0b}). 
For treating ${\wtilde {\cal S}}_{\lambda}^*$, much more complicated discussion than that in the case 
${\wtilde {\cal S}}_{\mu\nu}^*$ may be necessary. 
By calculating $\sum_{\lambda}s_{\lambda}{\wtilde {\cal S}}_{\lambda}^*$, we have 
the following relation: 
\beq\label{45}
\sum_{\lambda}s_{\lambda}{\wtilde {\cal S}}_{\lambda}^*=
-\frac{1}{4}\left({\wtilde S}_+\right)^2{\wtilde S}_-\ , \qquad
\sum_{\lambda}s_{\lambda}{\wtilde {\cal S}}_{\lambda}=
-\frac{1}{4}{\wtilde S}_+\left({\wtilde S}_-\right)^2\ .
\eeq
It is important to see that we have $\sum_{\lambda}s_{\lambda}{\wtilde {\cal S}}_{\lambda}^*\rket{m}
=\sum_{\lambda}s_{\lambda}{\wtilde {\cal S}}_{\lambda}\rket{m}=0$. 
Therefore, actually, we may regard $\sum_{\lambda}s_{\lambda}{\wtilde {\cal S}}_{\lambda}^*$ 
and $\sum_{\lambda}s_{\lambda}{\wtilde {\cal S}}_{\lambda}$ as vanishing 
operators in the minimum weight states. 
Thus, we obtain the minimum weight state in the form 
\beq
& &\rdket{p;(\lambda),q;(\mu\nu),rr_0;({\bar \alpha})}={\wtilde {\cal P}}^*(p;(\lambda),q;(\mu\nu))
\left({\wtilde R}_+\right)^{r+r_0}\rket{r;({\bar \alpha})}\ , 
\label{43}\\
& &{\wtilde {\cal P}}^*(p;(\lambda),q;(\mu\nu))=
\prod_{k=1}^p{\wtilde {\cal S}}_{\lambda_k}^*\prod_{j=1}^q{\wtilde {\cal S}}_{\mu_j\nu_j}^*\ . 
\label{44-1} 
\eeq
However, generally, it may be impossible to prove if the set of the states (\ref{43}) is 
orthogonal or not. 
If not, we must derive an appropriate orthogonal set from the set (\ref{43}), but, 
the general case may be impossible.

As the simplest example, we will treat the case of two-fermion minimum weight states. 
The state with the case ($r=1,\ q=0,\ p=0)$, $\rket{r=1;{\bar \alpha}_1{\bar \alpha}_2}$ and the 
states obtained under one- and two-time operation of ${\wtilde R}_+$ are given as 
\beq\label{46}
{\tilde c}_{{\bar \alpha}_1}^*{\tilde c}_{{\bar \alpha}_2}^*\rket{0}\ , \qquad
({\tilde c}_{{\alpha}_1}^*{\tilde c}_{{\bar \alpha}_2}^*+{\tilde c}_{{\bar \alpha}_1}^*{\tilde c}_{{\alpha}_2}^*)\rket{0}\ , \qquad
{\tilde c}_{{\alpha}_1}^*{\tilde c}_{{\alpha}_2}^*\rket{0}\ . 
\eeq
Of course, the seniority number is $n=2r=2$. 
Next, we treat other cases with $n=2$:\ ($r=0,\ q=1,\ p=0$) and 
$(r=0,\ q=0,\ p=1)$. 
For the former and the latter, the relation (\ref{43-0}) leads us to 
\beq
& &{\wtilde {\cal S}}_{\mu\nu}^*\rket{0}=\frac{1}{2}\Omega_0(2\Omega_0+1){\wtilde S}_{\mu\nu}^*\rket{0} \ , 
\label{47}\\
& &s_{\lambda}{\wtilde {\cal S}}_{\lambda}^*\rket{0}=\frac{1}{2}\Omega_0(2\Omega_0+1)
\left(s_{\lambda}{\wtilde S}_{\lambda}^*-\frac{1}{2\Omega_0}{\wtilde S}_+\right)\rket{0}\ . 
\label{48}
\eeq
Explicit form of the state (\ref{47}) is given as 
\beq\label{49}
{\wtilde {\cal S}}_{\mu\nu}^*\rket{0}
=\frac{1}{2}\Omega_0(2\Omega_0+1)({\tilde c}_{\mu}^*{\tilde c}_{\bar \nu}^*-{\tilde c}_{\bar \mu}^*{\tilde c}_{\nu}^*)\rket{0}\ . 
\eeq
If $\mu$ and $\nu$ read $\alpha_1$ and $\alpha_2$, respectively, we can notice the 
difference between the second of the states (\ref{46}) and the state (\ref{49}). 
This is the reply of the problem mentioned in (I). 
The relation (\ref{48}) leads us to 
\bsub\label{50}
\beq
& &\sum_{\lambda}s_{\lambda}{\wtilde {\cal S}}_{\lambda}^*\rket{0}=0\ , 
\label{50a}\\
& &\sum_{\lambda}z_{l,\lambda}s_{\lambda}{\wtilde {\cal S}}_{\lambda}^*\rket{0}
=\frac{1}{2}(2\Omega_0-1)\sum_{\lambda}z_{l,\lambda}s_{\lambda}{\wtilde S}_{\lambda}^*\rket{0}\ , 
\label{50b}\\
& &l=1,\ 2, \cdots ,\ 2\Omega_0-1\ . 
\nonumber
\eeq
\esub
Here, $\{z_{l,\lambda}\}$ denotes an orthogonal matrix in the $2\Omega_0$-dimension, which obey the 
condition 
\beq\label{51}
\sum_{\lambda}z_{l,\lambda}z_{l',\lambda}=\delta_{ll'}\ , \quad
z_{l=0,\lambda}=\left(\sqrt{2\Omega_0}\right)^{-1}. \ \ \  
(l,\ l'=0,\ 1,\ 2,\cdots ,\ 2\Omega_0-1)\quad
\eeq
An example of $z_{l,\lambda}$ is given in the form 
\setcounter{equation}{52}
\bsub
\beq
z_{l,\lambda}=\langle j\lambda j\!-\!\!\lambda\ket{l0}(-)^{j-\lambda}\ . \qquad
(j=\Omega_0-1/2)
\eeq
\esub
Here, $\langle j\lambda j\!-\!\lambda\ket{l0}$ denotes the Clebsch-Gordan coefficient and 
the relation $\langle j\lambda j\!-\!\lambda\ket{00}$\break
$=(\sqrt{2j+1})^{-1}(-)^{j-\lambda}$ is noted. 
From the relation (\ref{51}), the following is derived: 
\beq\label{52}
\sum_{\lambda}z_{l,\lambda}=0\ . \qquad (l=1,\ 2, \cdots ,\ 2\Omega_0-1)
\eeq
As is clear from the relation (\ref{50a}), the state $\sum_{\lambda}s_{\lambda}{\wtilde {\cal S}}_{\lambda}^*\rket{0}$ 
does not exist and the remaining states $\sum_{\lambda}z_{l,\lambda}s_{\lambda}{\wtilde {\cal S}}_{\lambda}^*\rket{0}\ 
(l=1,\ 2,\cdots ,\ 2\Omega_0-1)$ form an orthogonal set and 
correspond to the case $(r=0,\ q=0,\ p=1)$. 
The set $\{\sum_{\lambda} z_{l,\lambda}s_{\lambda}{\wtilde S}_{\lambda}^*\rket{0}\}$ forms an 
orthogonal set. 
As is shown in the relation (\ref{50}), they are expressed in terms of the original fermion operators.

Finally, we will contact with the four-fermion minimum weight states. 
Discussion on the minimum weight states in the aligned 
scheme $(r=2,\ q=0,\ p=0)$ is omitted. 
The states with $r=0$ are given in the following form: 
\bsub\label{53}
\beq
& &{\wtilde {\cal S}}_{\mu_2\nu_2}^*{\wtilde {\cal S}}_{\mu_1\nu_1}^*\rket{0}\ , \qquad (r=0,\ q=2,\ p=0)
\label{53a}\\
& &s_{\lambda}{\wtilde {\cal S}}_{\lambda}^*{\wtilde {\cal S}}_{\mu\nu}^*\rket{0}\ , \ \ \ \qquad (r=0,\ q=1,\ p=1)
\label{53b}\\
& &s_{\lambda_2}{\wtilde {\cal S}}_{\lambda_2}^*s_{\lambda_1}{\wtilde {\cal S}}_{\lambda_1}^*\rket{0}\ , \qquad (r=0,\ q=0,\ p=2)
\label{53c}
\eeq
\esub
With the use of the relation (\ref{42-0}), the state (\ref{53a}) is rewritten as
\beq\label{54}
{\wtilde {\cal S}}_{\mu_2\nu_2}^*{\wtilde {\cal S}}_{\mu_1\nu_1}^*\rket{0}
=\frac{1}{4}\Omega_0(\Omega_0-1)(2\Omega_0+1)(2\Omega_0-1){\wtilde S}_{\mu_2\nu_2}^*{\wtilde S}_{\mu_1\nu_1}^*\rket{0}\ . 
\eeq
The state (\ref{53b}) is expressed in the form 
\beq\label{54-1}
s_{\lambda}{\wtilde {\cal S}}_{\lambda}^*{\wtilde {\cal S}}_{\mu\nu}^*\rket{0}
&=&
\frac{1}{4}\Omega_0(\Omega_0-1)(2\Omega_0+1)\left(2\Omega_0-\frac{1}{2}\right)
\nonumber\\
& &\times \left[
s_{\lambda}{\wtilde S}_{\lambda}^*{\wtilde S}_{\mu\nu}^*
-\frac{1-\delta_{\lambda\mu}-\delta_{\lambda\nu}}{2(\Omega_0-1)}{\wtilde S}_+{\wtilde S}_{\mu\nu}\right]\rket{0}\ .
\eeq
We can easily verify the relation 
\beq\label{55}
\left(\sum_{\lambda}s_{\lambda}{\wtilde {\cal S}}_{\lambda}^*\right)
{\wtilde {\cal S}}_{\mu\nu}^*\rket{0}=0\ .
\eeq
This is a natural consequence of our treatment based on the relation ({\ref{45}). 
Further, we can prove the relation 
\beq\label{56}
& &
\left(\sum_{\lambda}z_{l,\lambda}^{(\mu\nu)}s_{\lambda}{\wtilde {\cal S}}_{\lambda}^*\right)
{\wtilde {\cal S}}_{\mu\nu}^*\rket{0}
\nonumber\\
&=&\frac{1}{4}\Omega_0(\Omega_0-1)(2\Omega_0+1)(2\Omega_0-1)
\left(\sum_{\lambda}z_{l,\lambda}^{(\mu\nu)}s_{\lambda}{\wtilde S}_{\lambda}^*\right)
{\wtilde S}_{\mu\nu}^*\rket{0}\ .
\eeq
Here, $z_{l,\lambda}^{(\mu\nu)}$ obeys the condition 
\bsub\label{57}
\beq
\sum_{\lambda}z_{l,\lambda}^{(\mu\nu)}
=z_{l,\mu}^{(\mu\nu)}+z_{l,\nu}^{(\mu\nu)}\ . 
\label{57a}
\eeq
For example, we have the expression 
\beq
z_{l,\lambda}^{(\mu\nu)}=z_{l,\lambda}+y_{l,\lambda}^{(\mu\nu)}\ , \qquad
\sum_{\lambda(\neq \mu,\nu)}y_{l,\lambda}^{(\mu\nu)}=
z_{l,\mu}+z_{l,\nu}\ . 
\label{57b}
\eeq
\esub
The relation (\ref{56}) tells us that the minimum weight state 
$(\sum_{\lambda}z_{l,\lambda}^{(\mu\nu)}s_{\lambda}{\wtilde {\cal S}}_{\lambda}^*){\wtilde {\cal S}}_{\mu\nu}^*\rket{0}$ 
can be explicitly expressed in terms of the original fermion operators. 
The state (\ref{53c}) is expressed as 
\beq\label{58}
s_{\lambda_2}{\wtilde {\cal S}}_{\lambda_2}^*s_{\lambda_1}{\wtilde {\cal S}}_{\lambda_1}^*\rket{0}
&=&
\frac{1}{4}\Omega_0(\Omega_0-1)(2\Omega_0+1)(2\Omega_0-1)
\nonumber\\
& &\times \biggl[s_{\lambda_2}{\wtilde {S}}_{\lambda_2}^*s_{\lambda_1}{\wtilde {S}}_{\lambda_1}^*\rket{0}
-\frac{1-\delta_{\lambda_1\lambda_2}}{2(\Omega_0-1)}{\wtilde S}_+
\left(s_{\lambda_1}{\wtilde S}_{\lambda_1}^*+s_{\lambda_2}{\wtilde S}_{\lambda_2}^*\right)\rket{0}
\nonumber\\
& &\ \ +\frac{1-\delta_{\lambda_1\lambda_2}}{2(\Omega_0-1)(2\Omega_0-1)}\left({\wtilde S}_+\right)^2\rket{0}
\biggl]\ . 
\eeq
The state (\ref{58}) is symmetric with respect to $\lambda_1$ and $\lambda_2$ and 
if $\lambda_1=\lambda_2$, this state vanishes. 
We can prove 
\bsub\label{59}
\beq
\left(\sum_{\lambda_2}s_{\lambda_2}{\wtilde {\cal S}}_{\lambda_2}^*\right)
s_{\lambda_1}{\wtilde {\cal S}}_{\lambda_1}^*\rket{0}
=s_{\lambda_2}{\wtilde {\cal S}}_{\lambda_2}^*\left(\sum_{\lambda_1}s_{\lambda_1}{\wtilde {\cal S}}_{\lambda_1}^*
\right)\rket{0}=0\ . 
\label{59a}
\eeq
Therefore, naturally we have 
\beq
\left(\sum_{\lambda_2}s_{\lambda_2}{\wtilde {\cal S}}_{\lambda_2}^*\right)
\left(\sum_{\lambda_1}s_{\lambda_1}{\wtilde {\cal S}}_{\lambda_1}^*\right)\rket{0}=0\ . 
\label{59b}
\eeq
\esub
The above is a natural consequence of our treatment shown in the relation (\ref{45}). 
Moreover, we have 
\beq\label{60}
& &\sum_{\lambda_1\lambda_2}z_{l_2l_1,\lambda_2\lambda_1}s_{\lambda_2}{\wtilde {\cal S}}_{\lambda_2}^*
s_{\lambda_1}{\wtilde {\cal S}}_{\lambda_1}^*\rket{0}\nonumber\\
&=&
\frac{1}{4}\Omega_0(\Omega_0-1)(2\Omega_0+1)(2\Omega_0-1)
\sum_{\lambda_1\lambda_2}z_{l_2l_1,\lambda_2\lambda_1}s_{\lambda_2}{\wtilde S}_{\lambda_2}^*
s_{\lambda_1}{\wtilde S}_{\lambda_1}^*\rket{0}\ . 
\eeq
Here, $z_{l_2l_1,\lambda_2\lambda_1}$ obeys the condition 
\beq\label{61}
z_{l_2l_1,\lambda_2\lambda_1}=z_{l_2l_1,\lambda_1\lambda_2}\ , \qquad
\sum_{\lambda_2}z_{l_2l_1,\lambda_2\lambda_1}=z_{l_2l_1,\lambda_1\lambda_1}\ . 
\eeq
For example, we have 
\beq\label{62}
& &z_{l_2l_1,\lambda_2\lambda_1}
=\frac{1}{2}(z_{l_2,\lambda_2}\cdot z_{l_1,\lambda_1}+z_{l_2,\lambda_1}\cdot z_{l_1,\lambda_2})
+x_{l_2l_1,\lambda_2\lambda_1}\ , 
\nonumber\\
& &\sum_{\lambda_2(\neq \lambda_1)}
x_{l_2l_1,\lambda_2\lambda_1}=z_{l_2,\lambda_1}\cdot z_{l_1,\lambda_1}\ . 
\eeq
Next, we consider the states with $r=1$: 
\bsub\label{63}
\beq
& &{\wtilde {\cal S}}_{\mu\nu}^*\rket{r=1;{\bar \alpha}_1{\bar \alpha_2}}\ , \qquad (r=1,\ q=1,\ p=0)
\label{63a}\\
& &s_{\lambda}{\wtilde {\cal S}}_{\lambda}^*\rket{r=1;{\bar \alpha}_1{\bar \alpha}_2}\ . 
\qquad (r=1,\ q=0,\ p=1)
\label{63b}
\eeq
\esub
Of course, the states obtained by the operations of ${\wtilde R}_+$ and 
$({\wtilde R}_+)^2$ are also 
the states with $r=1$. 
In the case of the form (\ref{63a}), it is simply 
given as 
\beq\label{64}
{\wtilde {\cal S}}_{\mu\nu}^*\rket{r=1;{\bar \alpha}_1{\bar \alpha}_2}
=\frac{1}{2}(\Omega_0-1)(2\Omega_0+1){\wtilde S}_{\mu\nu}^*\rket{r=1;{\bar \alpha}_1{\bar \alpha}_2}\ .
\eeq
It should be noted that the state (\ref{64}) obeys the rule in which 
$\mu$, $\nu$, $\alpha_1$ and $\alpha_2$ are all different of one another. 
The state (\ref{63b}) can be written as 
\beq\label{65}
s_{\lambda}{\wtilde {\cal S}}_{\lambda}^*\rket{r=1;{\bar \alpha}_1{\bar \alpha}_2}
&=&
\frac{1}{2}(\Omega_0-1)(2\Omega_0+1)\nonumber\\
& &\times
\left(s_{\lambda}{\wtilde S}_{\lambda}^*-\frac{1-\delta_{\lambda\alpha_1}-\delta_{\lambda\alpha_2}}{2(\Omega_0-1)}
{\wtilde S}_+\right)\rket{r=1;{\bar \alpha}_1{\bar \alpha}_2}\ . 
\eeq
Clearly, we have $\sum_{\lambda}s_{\lambda}{\wtilde {\cal S}}_{\lambda}^*\rket{r=1;{\bar \alpha}_1{\bar \alpha}_2}=0$. 
For other forms, we have 
\beq\label{66}
\sum_{\lambda}z'_{l,\lambda}s_{\lambda}{\wtilde {\cal S}}_{\lambda}^*\rket{r=1;{\bar \alpha}_1{\bar \alpha}_2}
=\frac{1}{2}(\Omega_0-1)(2\Omega_0+1)\sum_{\lambda}z'_{l,\lambda}s_{\lambda}{\wtilde S}_{\lambda}^*
\rket{r=1;{\bar \alpha}_1{\bar \alpha}_2}\ . 
\eeq
Here, $z'_{l,\lambda}$ obeys the condition 
\beq\label{67}
\sum_{\lambda}z'_{l,\lambda}=z'_{l,\alpha_1}+z'_{l,\alpha_2}\ .
\eeq
For example, we have 
\beq\label{68}
z'_{l,\lambda}=z_{l,\lambda}+w_{l,\lambda}\ , \qquad
\sum_{\lambda(\neq \alpha_1,\alpha_2)}w_{l,\lambda}=z_{l,\alpha_1}+z_{l,\alpha_2}\ . 
\eeq
If $w_{l,\lambda}$ is neglected, the operator $\sum_{\lambda}z'_{l,\lambda}s_{\lambda}{\wtilde S}_{\lambda}^*$ 
in the state (\ref{66}) becomes identical to the case (\ref{50b}). 
We omit the case of three-fermion states which are also simple. 
Of course, if we apply the present idea to concrete problems, 
we must check the orthogonality of the states (\ref{54}), (\ref{56}) and (\ref{60}). 
Through this task, $y_{l,\lambda}^{\mu\nu}$, 
$x_{l_2l_1, \lambda_2\lambda_2}$ and $w_{l,\lambda}$ may be determined. 
But, we do not intend to enter into this problem in this paper.

In this paper, together with (I), we presented an idea how to construct the minimum weight states in the 
$su(2)$-algebraic many-fermion model. 
The most important point can be found in introducing 
the auxiliary $su(2)$-algebra, which we called the $R$-spin. 
In addition to the aligned scheme developed in (I), 
in this paper, the scalar operators for the $R$-spin was introduced and we clarified the 
structure of the minimum weight states. 
The seniority number can be expressed in the form 
$n=2(r+q+p)$ and various formulae as functions of $2r$ appearing in (I) may be effective, 
if $2r$ is replaced with $n$.

\vspace{1cm}

One of the authors (Y.T.) 
is partially supported by the Grants-in-Aid of the Scientific Research 
(No.23540311) from the Ministry of Education, Culture, Sports, Science and 
Technology in Japan.



\begin{thebibliography}{99}
\bibitem{1}
Y. Tsue, C. Provid\^encia, J. da Provid\^encia and M. Yamamura, to appear in Prog. Theor. Phys. {\bf 127} (2012), No.1.
(arXive:1108:5816)
\end{thebibliography}
\end{document}